\documentstyle[aps,twocolumn,epsf,prb]{revtex}

\begin{document}

\draft

\wideabs{
\title
{Magnetic circular dichroism of x-ray absorption spectroscopy \\
at rare-earth {\textit{L}}$_{2,3}$ edges in RE$_{2}$Fe$_{14}$B compounds \\
(RE = La, Pr, Nd, Sm, Gd, Tb, Dy, Ho, Er, Tm, Yb, and Lu)
}%

\author{K. Fukui, H. Ogasawara, and A. Kotani}
\address{
Institute for Solid State Physics, University of Tokyo, \\
5-1-5 Kashiwanoha, Kashiwa, Chiba 277-8581, Japan
}%

\author{I. Harada and H. Maruyama}
\address{
Department of Physics, Faculty of Science, Okayama University, \\
3-1-1 Tsushima-naka, Okayama, Okayama 700-8530, Japan 
}%

\author{N. Kawamura}
\address{
The Institute of Physical and Chemical Research (RIKEN/SPring-8), \\
1-1-1 Kouto, Mikazuki, Sayo, Hyogo 679-5148, Japan
}%

\author{K. Kobayashi}
\address{
NEC Corporation, 34 Miyukigaoka, Tsukuba, Ibaraki 305-0801, Japan
}%

\author{J. Chaboy}
\address{
Instituto de Ciencia de Materiales de Arag\'{o}n, Zaragoza University, 
50009 Zaragoza, Spain
}%

\author{A. Marcelli}
\address{
INFN, Laboratori Nazionali di Frascati, C.P. 13, 00044 Frascati, 
Italy
}%

\date{\today}

\maketitle

\begin{abstract}
  Magnetic circular dichroism (MCD) in the x-ray absorption 
spectroscopy (XAS) at the $L_{2,3}$ edges for almost entire series 
of rare-earth (RE) elements in RE$_2$Fe$_{14}$B, 
is studied experimentally and theoretically.  
  By a quantitative comparison of the complicated MCD spectral shapes, 
we find that 
(i) the $4f$-$5d$ intra-atomic exchange interaction 
not only induces the spin and orbital polarization of the $5d$ 
states, which is vital for the MCD spectra of the electric dipole 
transition from the $2p$ core states to the empty $5d$ conduction 
band, but also it accompanies a contraction of the radial part of 
the $5d$ wave function depending on its spin and orbital state, 
which results in the enhancement of 
the $2p$-$5d$ dipole matrix element, 
(ii) there are cases where the spin 
polarization of the $5d$ states due to the hybridization 
with the spin polarized $3d$ states of surrounding irons plays important 
roles, and 
(iii) the electric quadrupole transition from the $2p$ core states 
to the magnetic valence $4f$ states is appreciable at the pre-edge region 
of the dipole spectrum. 
  Especially, our results evidence that it is important to include 
the enhancement effect of the dipole matrix element in the correct 
interpretation of the MCD spectra at the RE $L_{2,3}$ edges. 
\end{abstract}

}

\section{Introduction}

    In recent years, much work has been accumulated 
on the magnetic circular dichroism (MCD) 
in x-ray absorption spectroscopy (XAS) 
in various ferromagnetic and ferrimagnetic materials, 
showing the unique and powerful ability of the method to reveal 
detailed information on electronic and magnetic properties 
of a selected atom and even of a selected shell.  
This method has become one of the most powerful methods 
for the purpose, thanks to the recent technical progress 
in x-ray sources which provide strong intensity and high tunability 
as well as high quality of circularly polarized photons.

    In this paper, we treat the MCD spectra of XAS 
at the $L_{2,3}$ edges of almost entire series 
of rare-earth (RE) elements in metallic compounds, 
RE$_2$Fe$_{14}$B (Ref.~\onlinecite{JC-LMG-FB}). 
 Since the initial states, 
the $2p_{1/2}$ core state for the $L_{2}$ edge 
and the $2p_{3/2}$ core state for the $L_{3}$ edge which are split by 
the relatively large spin-orbit interaction, are well defined, 
we obtain from such a study detailed information on the final states, 
the $5d$ conduction band for the electric dipole (ED) transition 
and the valence $4f$ states for the electric quadrupole (EQ) transitions. 
 It is usually very hard to observe the information on the weak spin 
and orbital polarization of the $5d$ electron separately from others, 
although such information is important for understanding magnetic 
properties of these compounds since the $5d$ electron mediates 
the inter-atomic exchange process between $4f$-$4f$ electrons 
of RE's on different sites as well as $4f$-$3d$ electrons 
of RE and surrounding transition metal (TM) elements. 
 Thus, a detailed study of the MCD spectra is 
one of the most ideal methods for this purpose 
because of the selectivity of an atom and a shell mentioned before. 

    Unfortunately, however, there have been some problems 
in the interpretation of the MCD spectra at the $L_{2,3}$ edges 
of the RE elements: 
 A naive theory,~\cite{TJ-SI} 
which takes into account the polarization effect 
due to the $4f$-$5d$ exchange interaction, failed to explain 
the sign of the MCD integrated intensity of 
the experimental results,~\cite{FB-CG-SP} 
indicating the need for a more sophisticated interpretation 
of the spectra. 
 One of the reasons for this failure comes from the pliability 
of the extended $5d$ states, 
which are the final states of the ED transition. 
 The other comes from the ambiguity in estimating 
the contribution of the EQ transition to the $4f$ states. 
 Carra and co-workers~\cite{PC-BNH-BTT} have suggested 
that the EQ transition is appreciable at the pre-edge region 
of these edges since the $4f$ level is pulled down 
to this region due to the strong Coulomb interaction 
between the core hole and the $4f$ states. 
 And then, it has been tried to interpret the complicated structure 
as a consequence of the EQ contribution superimposed 
on the main ED component in a few RE-TM intermetallic 
compounds.~\cite{JCL-SWK-XDW,CG-ED-CB,KS-HM-KK,JCL-XDW-BNH} 
 However, these trials have not always succeeded 
in characterizing the EQ transitions, 
using, for instance, the angular dependence of the spectra. 
 To overcome this limitation, the systematic measurements 
for almost entire RE elements will be a great help. 

    The purpose of this paper is to measure the MCD spectra of XAS 
at the $L_{2,3}$ edges of almost entire series 
of RE elements in metallic compounds, RE$_2$Fe$_{14}$B 
and to make a theoretical analysis 
of these spectra in a systematic way. 
 In our theoretical model, 
it is essential to take into account the enhancement 
of the $2p$-$5d$ dipole matrix element due to the $4f$-$5d$ 
intra-atomic exchange interaction,~\cite{HM-IH-AK,MV-JBG-BTT} 
which depends not only on the $5d$ spin state but also 
on the $5d$ orbital state. 
 This effect is a consequence of the fact that 
according to the $4f$-$5d$ exchange interaction 
the radial part of the $5d$ wave function contracts 
and has a larger amplitude at the position of the $2p$ orbital. 
 We note here that this effect solves a serious 
discrepancy in sign of the MCD integrated intensity mentioned above. 

    In addition to this, we point out another important polarization 
effect of the $5d$ state due to the hybridization 
with the spin polarized $3d$ states of surrounding Fe ions. 
 It has been well known that the magnetic coupling 
between Fe $3d$ spin and RE $4f$ spin is always antiferromagnetic 
alignment via the conduction electrons, 
such as RE $5d$ electrons in the RE-TM intermetallic system, 
which makes them the ferro-magnetic (ferri-magnetic) compounds 
for the less-than-half (half-filled or more-than-half) RE ions. 
 At room temperature, Fe moments predominantly contribute to bulk 
magnetization, which has been verified 
by the Fe $K$-edge MCD spectrum.~\cite{JC-HM-LMG} 
 Since Fe constitutes the majority 
of the magnetic moment in RE$_2$Fe$_{14}$B, 
it is not difficult to realize that the effect is important 
in the MCD spectra at the RE $L_{2,3}$ edges. 
 In the case of La or Lu compound, this effect must 
dominate the spectra, since there is no effect from the $4f$ electrons. 
In fact, the MCD spectra in La$_2$Fe$_{14}$B and Lu$_2$Fe$_{14}$B 
observed are consistent with the result of the tight binding 
calculation for LaFe$_2$.~\cite{HO-KF-IH} 
 Then, as a first step, we take this effect into account 
phenomenologically: the experimental MCD spectrum 
of La$_2$Fe$_{14}$B is added complementally 
to the spectra of all other RE compounds, 
adjusting the amplitude and fixing the relative energy position. 
 This contribution makes sometimes the shape 
of the MCD spectra complicated. 

    On the other hand, the quadrupole contribution is 
also inevitable for a quantitative comparison. 
 Then, we calculate it using the Cowan's program~\cite{RDC} 
based on the atomic model, 
which is reasonable since the $4f$ electrons directly 
concerned with this process are well localized. 
 Many body effects are crucial in this process but the lifetime 
effect of the $2p$ core hole smears out a detailed structure 
of the spectra. 
 Superposing the electric quadrupole contribution on 
the dipole one, we reproduce the experimental spectra for the entire 
series of RE elements in RE$_2$Fe$_{14}$B. 

    In the next section, the experimental details and resulting spectra 
for XAS-MCD in RE$_2$Fe$_{14}$B are presented. 
 The theoretical model is described and 
the results are compared with the experimental ones 
in Sec.~\ref{sec:Results}. 
 In the last section, we summarize our results and give 
brief discussions on them. 

\section{Experimental Results}

    In this section, we describe the experimental conditions 
and present the observed spectra at the $L_{2,3}$ edges 
for almost entire series of RE elements 
in the metallic compounds, RE$_2$Fe$_{14}$B 
(RE = La, Pr, Nd, Sm, Gd, Tb, Dy, Ho, Er, Tm, Yb, 
and Lu).~\cite{JC-LMG-FB} 

    MCD measurements were made at room temperature 
in transmission mode using left-circularly polarized x-rays 
($+$ helicity) emitted from an elliptical multipole wiggler 
on the beamline 28B of the Photon Factory at KEK. 
 The beamline is composed of a fixed-exit double crystal monochromator 
equipped with Si(111) and Si(220) 
and two doubly focusing mirrors.~\cite{TI-AK-YS} 
 Powdered sample uniformly spread on the Scotch tape was used. 
 X-ray intensity was monitored using ionization chamber filled 
with N$_2$ gas before and after the sample: 
 Here we denote that $I_0$ is the intensity of incident beam 
while $I$ is that of transmitted beam. 
 Magnetic field of $0.6$ T was applied antiparallel or parallel 
to the direction of the incident x-ray wave vector, 
while the helicity was fixed.
 The sample plane was tilted $45^{\circ}$ away from the direction 
of the incident beam. 
 Degree of circular polarization $P_{\rm C}$ 
was estimated to be $0.35 \sim 0.6$ in the photon energy range studied. 
 Energy resolution has been assessed to be 
$\Delta E/E \sim 1.5 \times 10^{-4}$. 
 Energy dependence of absorption coefficient was recorded 
at an energy intervals of $1$ eV, and data were accumulated 
every $2$ seconds in order to minimize any time dependent drift, 
while the magnetic field was reversed twice for each energy point. 
 Such a measurement was repeated for $5$ to $15$ times.  

    XAS and MCD spectra are defined, respectively, as follows: 
\begin{equation}%
\mu t = \frac{1}{2}
         \left(
            {\mathrm{ln}} \frac{I_0}{I_{+}}
          + {\mathrm{ln}} \frac{I_0}{I_{-}}
         \right), 
\end{equation}
\begin{equation}
\Delta \mu t = {\mathrm{ln}} \frac{I_0}{I_{+}}
             - {\mathrm{ln}} \frac{I_0}{I_{-}}, 
\label{eq:MCD_definition}
\end{equation}%
where $I_{+} (I_{-})$ represents the intensity of transmitted x-ray 
with magnetization antiparallel (parallel) 
to the incident x-ray wave vector. 
 The XAS spectrum was normalized by absorption coefficient 
at higher energy side to obtain a thickness-independent spectrum, 
and the MCD spectrum was also subjected to the normalization and correction 
by tilting angle and degree of circular polarization.  

    The MCD spectra at the $L_{2,3}$ edges in RE$_2$Fe$_{14}$B 
(RE = La, Pr, Nd, Sm, Gd, Tb, Dy, Ho, Er, Tm, Yb, and Lu) 
thus obtained are shown with crosses 
in Fig.~1 
($L_{3}$) 
and Fig.~2 
($L_{2}$). 
 In these figures, the origin of the abscissa 
is taken to be the energy of the absorption edge $E_0$, 
which is determined as the energy 
at the first inflection point of XAS spectrum.

\section{Model and Calculated Results}
\label{sec:Results}

    In this section, we describe our model to calculate the XAS-MCD 
spectra at the $L_{2,3}$ edges of the RE elements, which are composed 
of two contributions; ED and EQ ones. 
 Then we discuss them separately in the following.

    First, we are concerned with the ED transition from the initial 
configuration $2p^6 4f^n 5d^1$ to the final configuration 
$2p^5 4f^n 5d^2$. 
 In the calculation of the XAS process, 
we make the following simplifications:  
 (i) The $5d$ states are so extended 
that they constitute the energy band 
having a simple semi-elliptic density of states. 
 The Coulomb and exchange interactions are ignored 
within the $5d$ electrons. 
 Furthermore, we assume a rectangular 
density of states following the above-mentioned semi-elliptic one, 
simulating higher energy $d$-symmetry states other 
than the $5d$ states of RE. 
 In this sense, we take a one-body picture for the $5d$ electrons.  
 (ii) On the other hand, the electron correlation within the $4f$ 
electrons is so strong that the Hund's-rule ground state is realized. 
 (iii) Then, the intra-atomic exchange interaction 
between the $4f$ and $5d$ electrons is considered as a mean-field 
while the intra-atomic Coulomb interaction between them 
is neglected since it does not affect the MCD spectra seriously. 

    Based on the assumptions (ii) and (iii), 
the energy of the $5d$ state specified by the $z$-component 
of the azimuthal quantum number, 
$m_d$, and that of the spin quantum number, $s_d$, is given by 
\begin{eqnarray}%
E_{d\mu}
 &\equiv&
     E(m_d,s_d) \nonumber \\
 &=& 
   - \sum_{k=1,3,5} \sum_{m_f,s_f} 
     \mid c^k(2m_d,3m_f) \mid^2 G^k \nonumber \\
 & & \times 
     n_{m_f,s_f}  \delta(s_d,s_f),
\end{eqnarray}%
where $d\mu$ denotes the combined indices of 
$m_d$ and $s_d$, $m_f$ and $s_f$ denote, respectively, 
the $z$-component of the azimuthal quantum number and 
that of the spin quantum number for $4f$ electrons, 
$c^k(lm_l,l'm_{l'})$ is proportional to the Clebsch-Gordan coefficients, 
$G^k$ ($k=1, 3, 5$) represent the $4f$-$5d$ Slater integrals 
which have been calculated 
using the Cowan's program~\cite{RDC} (Table~\ref{table:Hartree-Fock}), 
$n_{m_f,s_f}$ is the number of the $4f$ electrons 
in the state specified by $m_f$ and $s_f$, 
and $\delta(x,y)$ is the Kronecker delta function. 
 Here we note that the energy $E_{d\mu}$ depends on the number 
and their quantum numbers of the $4f$ electrons.

    As was mentioned above, the $5d$ states are assumed 
to form an energy band with the semi-elliptic density of states, 
$\rho_{d\mu}(\epsilon) = 2 \sqrt{W^{2} - (\epsilon - E_{d\mu})^{2}} 
                         / {\pi W^{2}}$, 
where $W$ denote the band width ($3.5$ eV), 
followed by the constant density of states 
having the $d$-symmetry in higher energies. 

    Denoting the core hole state as $pj(j=1/2, 3/2)$ 
and the photo-excited $5d$ state as $d\mu$, 
we calculate the absorption spectrum 
for the left- and right-circular polarized x-rays as 
\begin{eqnarray}%
F^{pj}_{\pm}(\omega)
 &=& \sum_{d\mu,j_{z}}
     \left| M_{pjj_{z},d\mu}^{\pm} \right|^{2}
     (1 - \alpha E_{d\mu}) \nonumber \\
 & & \times
    \int_{E_{\mathrm F}}^{W + E_{d\mu}}
    \!\!\!\!\!\!\!\!\!\!\!\!\!\!
    {\rm d}\varepsilon 
    \rho_{d\mu}(\varepsilon) 
    {\rm L}(\hbar\omega + E_{pj} - \varepsilon), 
\label{aa}
\end{eqnarray}%
and the corresponding MCD spectrum whose definition 
is consistent with the experimental condition as 
\begin{equation}%
\Delta F^{pj}(\omega) = F_{+}^{pj}(\omega) - F_{-}^{pj}(\omega). 
\end{equation}%
 Here $F^{pj}_{+}(\omega) (F^{pj}_{-}(\omega))$ represents 
the absorption spectrum due to the electric-dipole transition 
of an x-ray with positive (negative) helicity, 
whose matrix element is $M_{pjj_{z},d\mu}^{+} (M_{pjj_{z},d\mu}^{-})$ 
before taking into account the effect of the enhancement, 
and the factor, $(1 - \alpha E_{d\mu})$, parameterizes 
this enhancement effect depending on $d\mu$ (Ref.~\onlinecite{HM-IH-AK}). 
 This type of the enhancement factor is an extension 
of the actual observation in the band 
calculation~\cite{HK-XW-DBH} 
for Gd whose $4f$ electrons have only spin moments: 
 The $2p$-$5d$ ED matrix element 
for the $5d$ spin parallel to the $4f$ spin is 30\% larger than that 
for the $5d$ spin antiparallel to the $4f$ spin. 
 We fix the parameter value $\alpha$ to be $0.6$ for all RE elements, 
which yields the 30\% enhancement of the ED matrix element 
in the case of Gd metal. 
 The Fermi energy denoted by $E_{{\rm F}}$ is determined 
so that one $5d$ electron exists in the ground state.  
 $E_{pj}$ is the energy of the core state specified by $pj$ 
and L is the Lorentzian, ${\rm L}(x)=(\Gamma/\pi)/(x^2+\Gamma^2)$, 
where $2\Gamma$ denotes the spectral broadening due to the lifetime 
of the $2p$ core hole and is set to be $4.0$ eV. 
 The spectrum thus obtained is further convoluted 
with a Gaussian function of the width ($1.5$ eV) simulating 
the instrumental resolution to obtain the full spectrum. 

    As was mentioned in the introduction, 
we next consider the additional spin polarization effect 
of $5d$ states owing to the hybridization 
with the spin polarized $3d$ states of surrounding Fe ions. 
 This effect dominates the MCD spectra 
of La$_2$Fe$_{14}$B, since La has no $4f$ electron. 
 The MCD spectra of La$_2$Fe$_{14}$B exhibit the following 
characteristics: 
 (i) They have the intensity only in the energy 
range near the inflection point of XAS spectrum, 
 (ii) The MCD spectrum at the $L_3$ edge is mostly positive 
while the one at the $L_2$ edge is mostly negative. 
 These characteristics are hold in Lu$_2$Fe$_{14}$B and are supported 
by a tight-binding calculation for LaFe$_2$.~\cite{HO-KF-IH} 
 In our case, the inter-atomic hybridization 
as well as the intra-atomic exchange interaction contribute 
to the spin polarization of the $5d$ states. 
 In this paper, 
we take into account the hybridization effect phenomenologically: 
The MCD spectrum observed 
experimentally at each edge of La$_2$Fe$_{14}$B 
is added complementally to the calculated MCD spectrum 
(without the hybridization effect) of each RE compound, 
where the relative intensity of the two MCD spectra 
is treated as an adjustable parameter, 
while the energy position of the hybridization contribution 
is fixed with respect to the inflection point of each XAS spectrum. 

    On the other hand, for the EQ transition, 
we carry out the calculation based on the atomic model~\cite{RDC} 
with a reduction factor of $0.8$ for the Slater integrals 
since the states concerned with this transition, 
the initial state $2p^6 4f^n$ and the final state $2p^5 4f^{n+1}$, 
are well localized. 
 In our calculation, we ignore the existence 
of a $5d$ electron in the ground state for simplicity.  
 Although the interactions between a photo-excited $4f$ electron and 
the core hole as well as other $4f$ electrons are crucial in this process, 
the lifetime effect of the $2p$ core hole smears out a detailed 
structure of the spectra.
 Then, we superpose the ED spectrum and the EQ spectrum, 
obtaining the total one; the relative energies of these spectra 
have been estimated by the atomic Hartree-Fock calculations~\cite{RDC} 
while their intensity ratio is regarded as a fitting parameter.  
 It is  to be noted that the MCD of EQ transition 
has a strong angle dependence,~\cite{PC-BNH-BTT,MV-RB} 
and we have fixed the angle $\theta$ 
between the $z$-axis (quantization axis) 
and the direction of the incident x-ray wave vector to be 
$45^{\circ}$, so as to fit to the experimental geometry.

    Now, we are ready to compare the total MCD spectra, 
which are the superposition of the ED and EQ contributions, 
with the experimental spectra for a series of RE elements 
in RE$_2$Fe$_{14}$B 
in Figs.~1 and 2. 
 Here, the calculated spectra of both XAS and MCD are adjusted 
so that the calculated maximum intensities of these spectra 
coincide with the corresponding experimental ones.
 The agreement is quite good for all RE elements. 
 In order to see the spectral structure in more detail, 
we present the results for 
Nd$_2$Fe$_{14}$B (Fig.~3), 
Gd$_2$Fe$_{14}$B (Fig.~4), and 
Er$_2$Fe$_{14}$B (Fig.~5), 
which are typical examples of the light RE (Nd), 
of the half-filled case (Gd), and of the heavy RE (Er). 
 In these figures, we use the values of the energy difference 
$|\Delta E|$ between the ED and EQ transitions 
for the trivalent RE ions tabulated in Table~\ref{table:Hartree-Fock}, 
which have been calculated based on the atomic model.~\cite{RDC} 

    In these spectra, we see the following: 
The ED contribution is decomposed into two components, 
 (i) the effect due to the $4f$-$5d$ exchange interaction  
(the dashed curve in Figs.~3-5) and 
 (ii) the effect of hybridization between the RE $5d$ and Fe $3d$ bands 
(the dotted curve). 
 In the contribution (i), 
the enhancement effect of the dipole matrix element is dominant 
compared with the magnetic polarization effect of the $5d$ band 
(see Ref.~\onlinecite{HM-IH-AK}). 
 On the other hand, the EQ contribution (the chain curve) is appreciable 
in the MCD spectra only at the $L_{3}$ pre-edge region. 
 For the ED contribution, 
there are cases where the signs of the MCD spectra  
due to the enhancement and the hybridization effects are opposite.
 This situation is clearly found at Er 
(the left panel of Fig.~5) or 
Tm $L_{3}$ edge.
 The enhancement effect is dominant 
in the $L_{2}$ edge of Pr, Nd, Sm, Gd, and 
in the $L_{3}$ edge of Gd, Tb, Dy, Ho, Er, Tm. 
 The MCD signals have the peak near the white line 
but slightly below the maximum of XAS spectra.
 On the other hand, the hybridization effect is well recognized 
not only in the La and Lu $L_{2,3}$ edges but also 
in the $L_{2}$ edge of Er, Tm, Yb, and 
in the $L_{3}$ edge of Pr, Nd, Sm, Yb, 
since the contribution of the $4f$-$5d$ exchange effect 
is originally small to the MCD integrated intensities 
in these cases.~\cite{HM-IH-AK} 
 The EQ contribution is recognized in the MCD spectra at the
 $L_{3}$ pre-edge of several RE's and has an opposite sign 
to the ED one by the enhancement effect. 
 In the $L_{2}$ edge of heavy RE elements, the EQ signal is very small. 
 We will discuss this in the next section.

\section{Summary and Discussions}
\label{sec:summary}

    We found that the MCD spectra observed at the $L_{2,3}$ edges 
for almost entire series of RE elements in RE$_2$Fe$_{14}$B are well 
reproduced if we take into account 
for the ED transition 
 (i) the enhancement of the $2p$-$5d$ ED transition matrix element 
caused by the $4f$-$5d$ intra-atomic exchange interaction and 
 (ii) the spin and orbital polarization of the $5d$ states 
due to this interaction as well as 
the hybridization with the spin polarized $3d$ states 
of the surrounding Fe ions, and 
 (iii) the EQ transition at the pre-edge region. 
 The fine structure of the density of states of the $5d$ conduction 
band is considered to be smeared by the large spectral width 
due to the lifetime of the deep $2p$ core hole, 
although the hybridization effect might reflect more or less the band 
structure of the $5d$ and $3d$ states. 

    In addition, we note that the $5d$ electron in the initial state 
introduces much varieties to XAS-MCD spectra:  
 The $5d$ electron in the initial state occupying the low energy state 
blocks $2p$-$5d$ dipole excitation to the lower energy states,
while the enhancement factor becomes larger for the lower energy states 
and eventually overcomes the blocking effect in our calculation. 
 If we consider only the $4f$-$5d$ intra-atomic exchange interaction 
(including the enhancement effect), we can say the following: 
 In the case of light RE elements, 
the positive MCD spectra in lower energy range 
are enhanced significantly and this results 
in the large positive integrated intensity of MCD at the $L_2$ edge.
 For the half-filled case (Gd), 
the enhancement effect causes the large positive and negative MCD 
in the $L_2$ and $L_3$ edges, respectively (see Fig.~4). 
 On the other hand, in heavy RE elements, 
the MCD spectra at the $L_3$ edge show a similar 
behavior to that in the $L_2$ edge of light RE elements 
except for the reversed signs. 
 The intensity of MCD at the $L_3$ edge of light RE elements 
and the $L_2$ edge of heavy RE elements is small, 
because it is determined  by a delicate balance 
of the positive and negative contributions. 
 Then the hybridization effect with the $3d$ electrons 
becomes very important, as seen in Fig.~3 ($L_3$) 
and Fig.~5 ($L_2$). 

    Now, let us consider the hybridization effect qualitatively. 
 It has been mentioned that this effect occurs in the restricted 
energy range near the inflection point of XAS, 
which is reasonable since the $5d$ energy band 
only near the Fermi energy is strongly affected 
by the hybridization effect. 
 Also it is interesting to point out 
that this effect yields the sign of the MCD spectrum, 
being consistent with what we expect in the case where the spin 
of the $5d$ electron is antiparallel to that of the Fe $3d$ electrons. 
 Our preliminary tight-binding calculation 
for MgCu$_2$-type Laves phase compounds 
supports these characteristics.~\cite{HO-KF-IH} 
 Since this effect is free from the enhancement of the ED matrix element, 
this gives rise to the contribution 
with the different signal to the MCD spectra than 
the contribution from the $4f$-$5d$ intra-atomic exchange interaction. 
 These observations are quite consistent with experimental results. 

    We now turn to the case of the EQ contribution. 
 In this case, the $4f$ electrons play a direct role 
since the $2p$ electron is excited to the $4f$ state. 
 The most important interaction is surely 
the Coulomb and the exchange interactions among $4f$ electrons, 
resulting in forming the multiplet structure which can be calculated 
with the aid of the atomic Cowan's program.~\cite{RDC} 
 Each spectrum of light RE elements consists of two groups, 
which correspond to the final states having a parallel spin 
and an antiparallel spin of the photo-excited electron 
to the $4f$ spin in the initial state. 
 In the case of heavy RE elements, 
there is only one kind of spin in unoccupied $4f$ states so that 
the structure in the spectra is not simply divided into two groups. 
 The fine structures discussed above is unfortunately smeared 
by the short lifetime of the deep $2p$ core hole and the spectra 
in fact show a very simple Lorentzian-like shape. 
 In the experimental MCD spectra, the EQ transition is apparent 
at the $L_3$ edge, while the traces of the EQ transition are not found 
at the $L_2$ edge.
 This is consistent with our calculations. 
 It can be shown from the atomic calculation~\cite{KF} 
that the MCD intensity of the EQ transition 
at the $L_2$ edge is much smaller, 
for most of RE elements, than that at the $L_3$ edge 
for $\theta = 45^{\circ}$. 
 If the angle $\theta$ is changed to $0^{\circ}$, 
according to the theoretical calculation, 
the MCD intensity of the EQ transition 
at the $L_2$ edge should be increased for Pr, Nd, and Tb. 
 It is desirable that this is confirmed 
by experimental observations in future. 

    The x-ray emission spectroscopy (XES), 
in which the final state of XAS is the intermediate state and 
then the $3d$ core electron fills the $2p$ core hole with emitting x-ray, 
is another tool to identify the EQ contribution experimentally 
since the broadening of the spectra is reduced by virtue of the relatively 
long lifetime of the $3d$ core hole to the $2p$ core hole.~\cite{FB-HMT-LS} 
 Observation of XES at the $L_2$ edge is desired. 

    In conclusion, we have confirmed the important role of the $4f$-$5d$ 
exchange interaction, which leads to the spin and orbital dependent 
enhancement of the $2p$-$5d$ ED matrix element in addition to the usual 
polarization effect in XAS-MCD. 
 We find that, in some cases, the hybridization of the $5d$ states 
with the spin polarized $3d$ states of the surrounding Fe ions 
plays a vital role.  
 Furthermore, the EQ transition is appreciable, 
especially, at the $L_3$ edge. 

\begin{acknowledgments}
    The authors thank Dr. T. Iwazumi of KEK-PF 
for his technical suggestions and supports. 
 The experiments in this work have been performed with the approval 
of Photon Factory Program Advisory Committee (Proposal No.94-G161) 
while the computation has been done using the facilities 
of the Supercomputer Center, ISSP, University of Tokyo.
\end{acknowledgments}

\onecolumn

\begin{figure*}[tbhp]
\begin{center}
  \epsfxsize=.8\textwidth
  \epsfbox{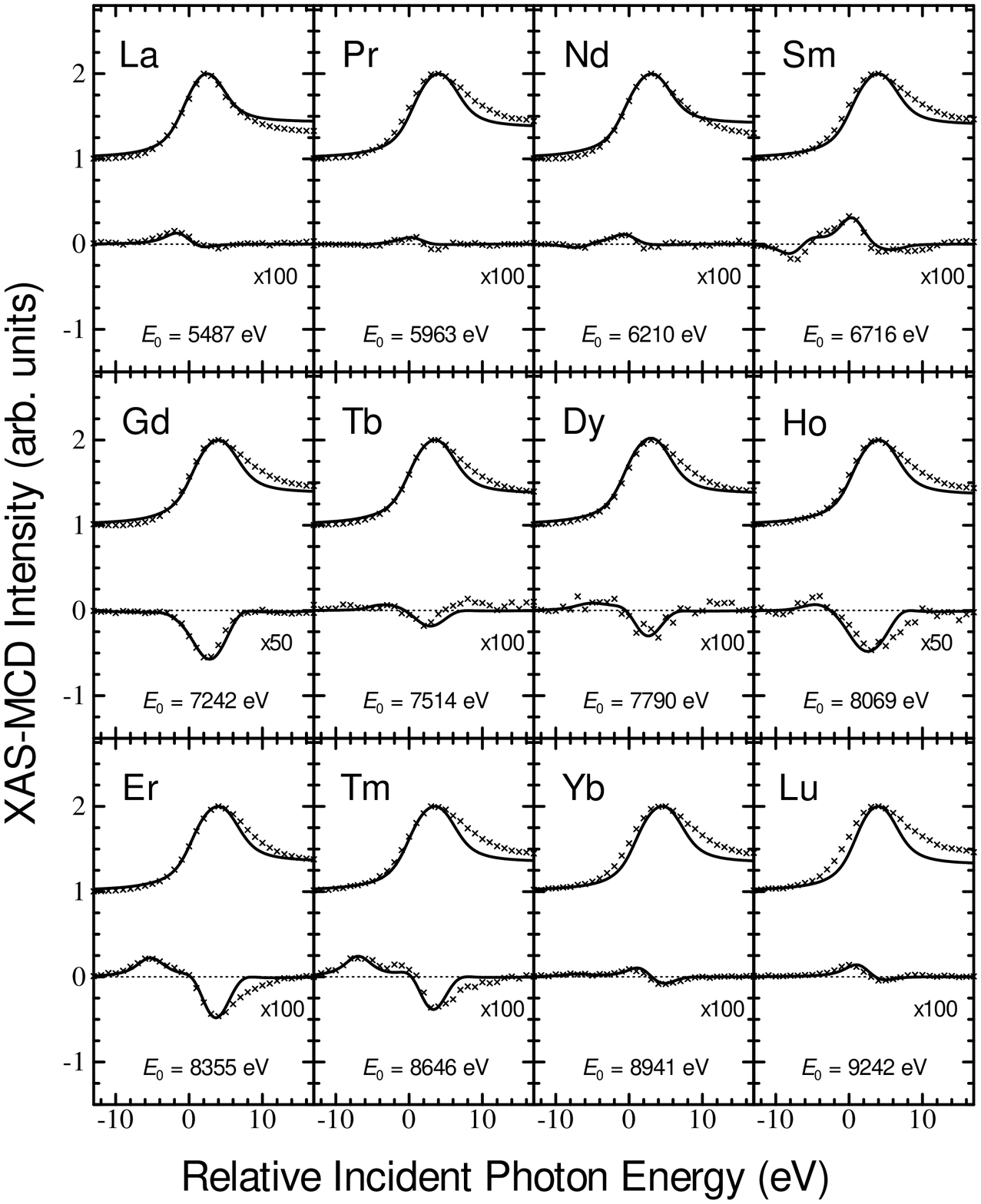}
\caption{The observed XAS and MCD spectra (crosses) 
at the RE $L_{3}$ edge in RE$_2$Fe$_{14}$B compounds (RE: rare-earth). 
  The normalized XAS spectra have been offset by unit for clarity. 
  The MCD intensities are also normalized by the peaks 
of the XAS spectra and are multiplied by a factor noted in each panel. 
  The origin of the energy axes represents the inflection point ($E_{0}$) 
of each XAS spectrum. 
  The solid curves are calculated spectra of $L_{3}$ XAS and MCD spectra 
for RE$^{3+}$ but the curves of MCD for La and Lu 
are not the calculated ones (see text). } 
\end{center}
\label{fig:L3XASMCD}
\end{figure*}

\onecolumn

\begin{figure*}[tbhp]
\begin{center}
  \epsfxsize=.8\textwidth
  \epsfbox{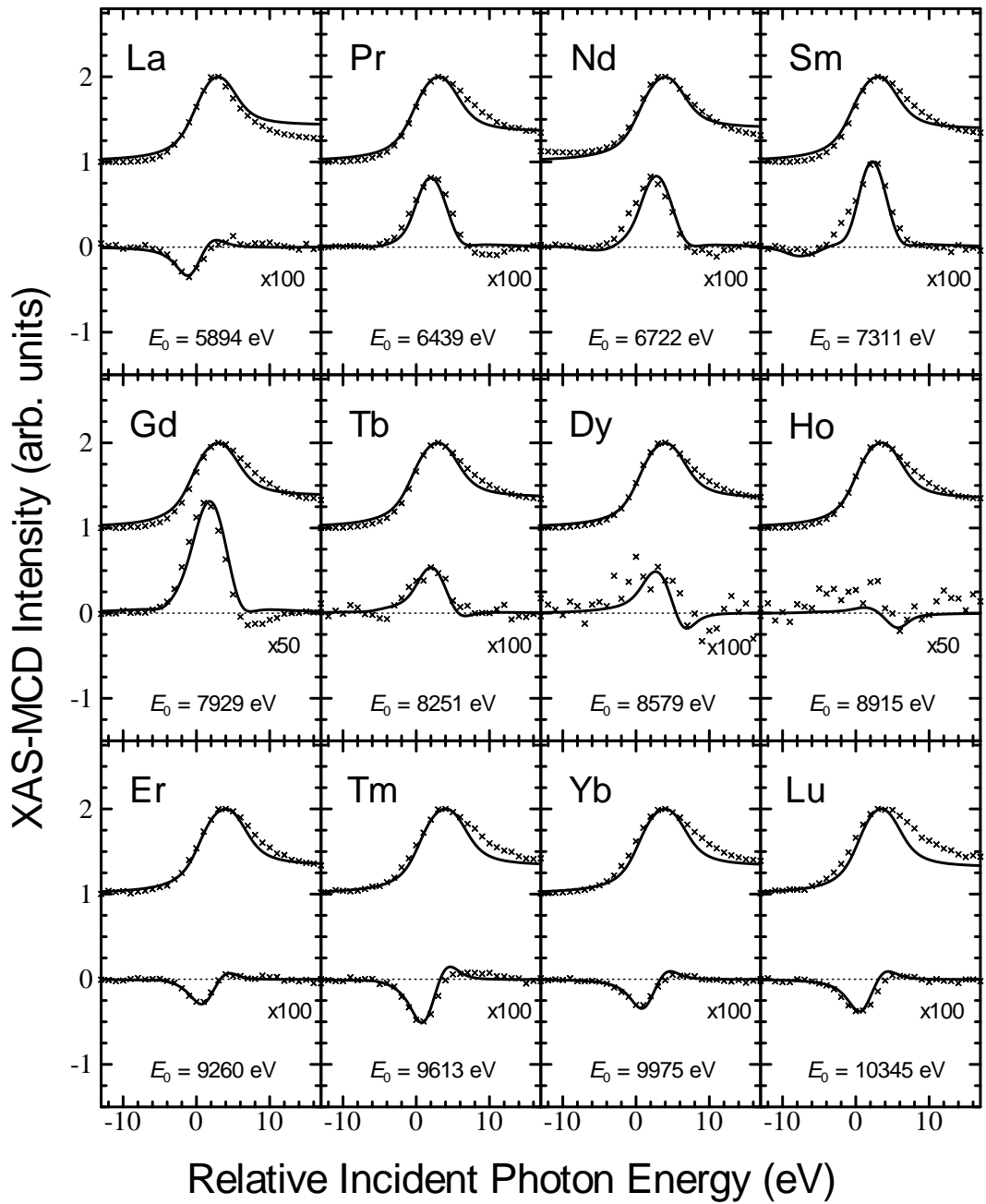}
\caption{The same as Fig.~1 
but for RE $L_{2}$ edge.} 
\end{center}
\label{fig:L2XASMCD}
\end{figure*}

\twocolumn

\begin{figure}[tbhp]
\begin{center}
  \epsfxsize=0.9\linewidth
  \epsfbox{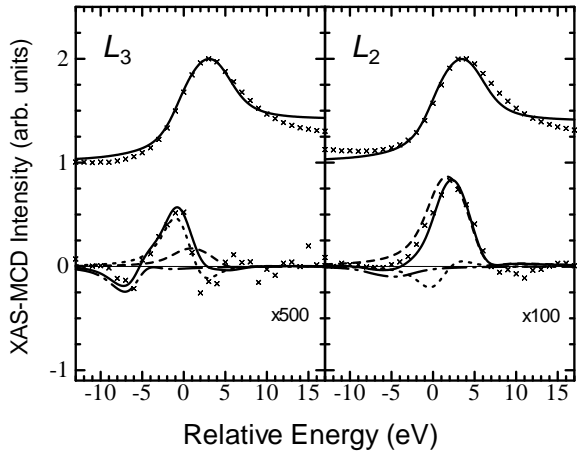}
\caption{$L_{2,3}$ XAS and MCD for Nd$^{3+}$ in Nd$_2$Fe$_{14}$B, 
as a typical example of light RE elements. 
  The origin of the energy is chosen as $E_{0}$. 
  The solid curve is the calculated result, 
which consists of the ED contribution 
due to the $4f$-$5d$ exchange interaction 
(the dashed curve), the ED contribution due to the hybridization 
with the Fe $3d$ electrons (the dotted curve) and the EQ contribution 
(the chain curve), while the crosses represent 
the experimental results. } 
\end{center}
\label{fig:Nd_XAS-MCD}
\end{figure}

\begin{figure}[tbhp]
\begin{center}
  \epsfxsize=0.9\linewidth
  \epsfbox{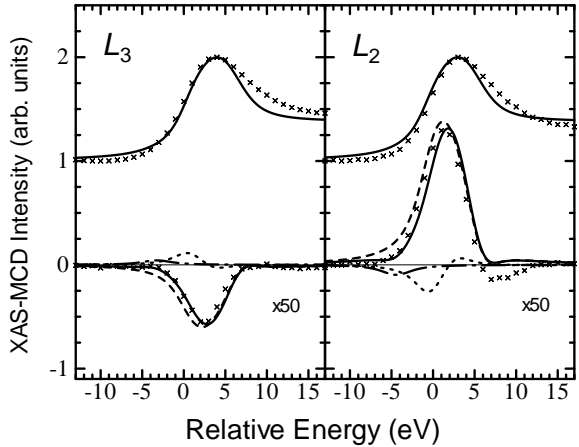}
\end{center}
\caption{$L_{2,3}$ XAS and MCD for Gd$^{3+}$ as a typical example 
of having no $4f$ orbital moment. 
  Others are the same as in Fig.~3. 
} 
\label{fig:Gd_XAS-MCD}
\end{figure}

\begin{figure}[tbhp]
\begin{center}
  \epsfxsize=0.9\linewidth
  \epsfbox{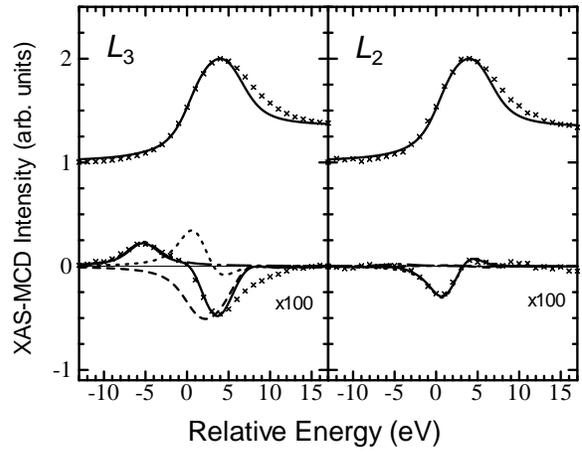}
\end{center}
\caption{$L_{2,3}$ XAS and MCD for Er$^{3+}$ 
as a typical example of heavy RE elements. 
  Others are the same as in Fig.~3. 
} 
\label{fig:Er_XAS-MCD}
\end{figure}

\begin{table}[bthp]
\begin{center}
 \catcode`?=\active \def?{\phantom{0}}
\begin{tabular}[b]{ccccc} 
       & $|\Delta E|$ &  $G^{1}$  &  $G^{3}$  &  $G^{5}$  \\
\hline
Pr$^{3+}$ & $ ?7.86 $ & $ 1.293 $ & $ 1.021 $ & $ 0.773 $  \\
Nd$^{3+}$ & $ ?8.41 $ & $ 1.205 $ & $ 0.992 $ & $ 0.752 $  \\
Sm$^{3+}$ & $ ?9.30 $ & $ 1.184 $ & $ 0.946 $ & $ 0.718 $  \\
Gd$^{3+}$ & $ ?9.97 $ & $ 1.134 $ & $ 0.909 $ & $ 0.690 $  \\
Tb$^{3+}$ & $ 10.23 $ & $ 1.114 $ & $ 0.893 $ & $ 0.677 $  \\
Dy$^{3+}$ & $ 10.45 $ & $ 1.095 $ & $ 0.877 $ & $ 0.665 $  \\
Ho$^{3+}$ & $ 10.62 $ & $ 1.078 $ & $ 0.862 $ & $ 0.654 $  \\
Er$^{3+}$ & $ 10.76 $ & $ 1.062 $ & $ 0.849 $ & $ 0.643 $  \\
Tm$^{3+}$ & $ 10.86 $ & $ 1.046 $ & $ 0.835 $ & $ 0.633 $  \\
Yb$^{3+}$ & $ 10.93 $ & $ 1.032 $ & $ 0.822 $ & $ 0.622 $
\end{tabular}
\end{center}
\caption{The atomic values of the Slater integrals $G^{k}$ (eV) 
for the $4f$-$5d$ exchange interaction 
and the energy difference $|\Delta E|$ (eV) 
between the term-averaged energies of the $2p$-$5d$ ED 
and $2p$-$4f$ EQ transition in RE$^{3+}$ ion. 
 These are calculated using the atomic Cowan's program 
based on the Hartree-Fock method. 
}
\label{table:Hartree-Fock}
\end{table}%


\begin{references}
\bibitem{JC-LMG-FB}
A short report of the experimental data was given in the 
{\it Proceedings of the 9th International Conference 
on X-ray Absorption Fine Structure}: 
J. Chaboy, L. M. Garc\'{\i}a, F. Bartolom\'{e}, 
J. Bartolom\'{e}, H. Maruyama, K. Kobayashi, N. Kawamura, 
A. Marcelli, and L. Bozukov, 
J. Phys. IV FRANCE {\bf 7}, C2-449 (1997). 
\bibitem{TJ-SI}
T. Jo and S. Imada, 
J. Phys. Soc. Jpn. {\bf 62}, 3721 (1993). 
\bibitem{FB-CG-SP}
See, for example, F. Baudelet, Ch. Giorgetti, S. Pizzini, 
Ch. Brouder, E. Dartyge, A. Fontaine, J. P. Kappler, and G. Krill, 
J. Electron Spectrosc. Relat. Phenom. {\bf 62}, 153 (1993). 
\bibitem{PC-BNH-BTT}
P. Carra, B. N. Harmon, B. T. Thole, M. Altarelli, and G. A. Sawatzky,
Phys. Rev. Lett. {\bf 66}, 2495 (1991). 
\bibitem{JCL-SWK-XDW}
J. C. Lang, S. W. Kycia, X. D. Wang, B. N. Harmon, A. I. Goldman, 
D. J. Branagan, R. W. McCallum, and K. D. Finkelstein, 
Phys. Rev. B{\bf 46}, 5298 (1992). 
\bibitem{CG-ED-CB}
Ch. Giorgetti, E. Dartyge, Ch. Brouder, F. Baudelet, C. Meyer, 
S. Pizzini, A. Fontaine, and R. M. Gal\'{e}ra, 
Phys. Rev. Lett. {\bf 75}, 3186 (1995). 
\bibitem{KS-HM-KK}
K. Shimomi, H. Maruyama, K. Kobayashi, A. Koizumi, H. Yamazaki, 
and T. Iwazumi, Jpn. J. Appl. Phys. {\bf 32-2}, 314 (1993). 
\bibitem{JCL-XDW-BNH}
J. C. Lang, X. D. Wang, B. N. Harmon, A. I. Goldman, K. W. Dennis, 
R. W. McCallum, and K. D. Finkelstein, 
Phys. Rev. B{\bf 50}, 13805 (1994). 
\bibitem{HM-IH-AK}
H. Matsuyama, I. Harada, and A. Kotani, 
J. Phys. Soc. Jpn. {\bf 66}, 337 (1997). 
\bibitem{MV-JBG-BTT}
M. van Veenendaal, J. B. Goedkoop, and B. T. Thole, 
Phys. Rev. Lett. {\bf 78}, 1162 (1997). 
\bibitem{JC-HM-LMG}
J. Chaboy, H, Maruyama, L. M. Garc\'{\i}a, F. Bartolom\'{e}, 
K. Kobayashi, N. Kawamura, A. Marcelli, and L. Bozukov, 
Phys. Rev. B{\bf 54}, R15637 (1996). 
\bibitem{HO-KF-IH}
H. Ogasawara, K. Fukui, I. Harada, and A. Kotani, 
Meeting Abstracts of Phys. Soc. Jpn. {\bf 55}, Issue 2, 
Part 4, 603 (2000).  
\bibitem{RDC}
R. D. Cowan, {\it The Theory of Atomic Structure and Spectra} 
(University of California Press, Berkeley, 1981). 
\bibitem{TI-AK-YS}
T. Iwazumi, A. Koyama, and Y. Sakurai, 
Rev. Sci. Instrum. {\bf 66}, 1691 (1995). 
\bibitem{HK-XW-DBH}
H. K\"onig, X. Wang, B. N. Harmon, and P. Carra, 
J. Appl. Phys. {\bf 76}, 6474 (1994). 
\bibitem{MV-RB}
M. van Veenendaal and R. Benoist, 
Phys. Rev. B{\bf 58}, 3741 (1998). 
\bibitem{KF}
K. Fukui, doctoral thesis, Okayama University, 2000. 
\bibitem{FB-HMT-LS}
F. Bartolom\'{e}, J. M. Tonnerre, L. S\`{e}ve, D. Raoux, J. Chaboy, 
L. M. Garc\'{\i}a, M. Krisch, and C. C. Kao, 
Phys. Rev. Lett. {\bf 79}, 3775 (1997). 
\end{references}
\end{document}